\documentclass[fleqn,usenatbib]{mnras}

\usepackage{newtxtext,newtxmath}
\usepackage[T1]{fontenc}

\DeclareRobustCommand{\VAN}[3]{#2}
\let\VANthebibliography\thebibliography
\def\thebibliography{\DeclareRobustCommand{\VAN}[3]{##3}\VANthebibliography}

\usepackage{graphicx}	% Including figure files
\usepackage{amsmath}	% Advanced maths commands

\usepackage{paralist}
\usepackage{verbatim}

\usepackage{rotating}
\usepackage{float}
\usepackage{xcolor, soul}
\usepackage{hyperref}

\usepackage{subfigure}
\usepackage[normalem]{ulem}
\usepackage[bottom]{footmisc}
\usepackage{adjustbox}
\usepackage{graphics}

\usepackage{natbib}
\usepackage{textcomp}
\usepackage{enumitem}

\DeclareRobustCommand{\okina}{%
  \raisebox{\dimexpr\fontcharht\font`A-\height}{%
    \scalebox{0.8}{`}%
  }%
}

\newcommand{\msun}{{\rm M}_\odot}

\title[Precision constraints on stellar physics]{Precision constraints on stellar physics from main sequence detached eclipsing binaries}

\author[Mitchell T. Dennis et al.]{
Mitchell T. Dennis,$^{1}$\thanks{E-mail: mtde226@hawaii.edu}
Harry Desmond$^{2}$,
and Jeremy Sakstein$^{3}$
\\
$^{1}$Institute for Astronomy, University of Hawaii at M\=anoa, 2680 Woodlawn Drive, Honolulu, HI 96822, USA\\
$^{2}$Institute of Cosmology and Gravitation, University of Portsmouth, Dennis Sciama Building, Portsmouth PO1 3FX, United Kingdom\\
$^{3}$Department of Physics \& Astronomy, University of Hawaii at M\=anoa, Watanabe Hall 2505 Correa Road, Honolulu, HI 96822 USA\\
}

\date{Accepted XXX. Received YYY; in original form ZZZ}

\pubyear{\the\year{}}

\begin{document}
\label{firstpage}
\pagerange{\pageref{firstpage}--\pageref{lastpage}}
\maketitle


\begin{abstract}
    We present a Bayesian framework to constrain {mass ($M$), metallicity ($Z$), stellar age ($\tau$), and the convective mixing length parameter ($\alpha_{\rm MLT}$)} in main-sequence (MS) detached eclipsing binaries (DEBs). These systems provide precise values of stellar mass and radius, offering stringent tests of stellar evolution models. We combine these with broadband magnitudes in the $B$ and $V$ bands and Gaussian priors on spectroscopic mass and metallicity, and perform Markov Chain Monte Carlo inference using a fast machine-learning surrogate for one-dimensional stellar evolution models computed with Modules for Experiments in Stellar Astrophysics. To make this approach computationally feasible, we implement an active learning strategy that adaptively selects new stellar models to evaluate, concentrating training data in regions of parameter space where the surrogate is most uncertain. Applying this framework to 38 stars in DEB systems, we recover ages more precise than previous isochrone-based determinations and obtain bounds on $\alpha_{\rm MLT}$ for a subset of lower-mass stars ($M \lesssim 1.5 M_\odot$), where convective envelopes provide sensitivity to the mixing length parameter. For several stars, the inferred $\alpha_{\rm MLT}$ values lie below the Solar-calibrated value, supporting previous indications that a universal mixing length parameter may not adequately describe convection across the main sequence. The active learning methods developed here provide a scalable route to Bayesian inference with stellar evolution models, with clear applications to additional stellar physics parameters and other precisely characterized stellar systems.
\end{abstract}


%%%%%%%%%%%%%%%%%%%%%%%%%%%%%%%%%%%%%%%%%%%%%%%%%%

%%%%%%%%%%%%%%%%% BODY OF PAPER %%%%%%%%%%%%%%%%%%

\section{Introduction}

Stellar ages and stellar evolution models underpin much of astrophysics, from dating star clusters and calibrating the cosmic distance ladder to interpreting stellar populations in galaxies. Central to these models is the treatment of convection, which in one-dimensional stellar evolution codes is typically described by mixing length theory \citep{Bohm1958} with a free parameter, $\alpha_{\rm MLT}$, calibrated to reproduce the present-day Sun.~$\alpha_{MLT}$ represents the efficiency of stellar convection, with higher values yielding higher $T_{\rm eff}$, smaller radii, and lower luminosity in stellar models. Despite its widespread use, there is no fundamental reason to expect this Solar-calibrated value to apply universally across different stellar masses, metallicities, and evolutionary states. Indeed, both theoretical and observational studies have suggested that $\alpha_{\rm MLT}$ may vary  with stellar properties such as effective temperature or surface gravity \citep{Magic2015, Tayar2017, Song2020}. If $\alpha_{\rm MLT}$ varies systematically, this would directly affect inferred stellar ages, radii, and luminosities, with consequences for a broad range of astrophysical applications. Testing the assumption of a universal mixing length parameter, and quantifying possible deviations, therefore represents a key step toward improving the physical fidelity of stellar evolution models.

An important opportunity to test mixing length parameter universality is provided by main-sequence detached eclipsing binaries (MS DEBs). These systems consist of two hydrogen-burning stars that are sufficiently separated to avoid mass transfer while remaining aligned such that they eclipse one another. Observations of their light curves and radial velocities yield precise, largely model-independent measurements of stellar masses and radii, along with broadband magnitudes \citep{Paczynski1997, Stassun2009}. Such measurements provide strong constraints on stellar structure models and therefore offer a powerful way to infer otherwise poorly determined parameters such as stellar age and $\alpha_{\rm MLT}$. In this work we develop a Bayesian framework to leverage these measurements and constrain both stellar age and $\alpha_{\rm MLT}$ in MS DEB stars.

Historically, attempts to constrain the age and mixing length parameter used isochrone fitting, which uses $\chi$-squared minimization to find the best fit of the stellar tracks overlaid on a color--magnitude diagram. However, $\chi$-squared analyses rely on Gaussian error propagation which fails when the likelihood is asymmetric or otherwise non-Gaussian, struggles to capture complex relationships between parameters, and typically does not incorporate prior information. Conversely, Bayesian inference through Markov Chain Monte Carlo (MCMC) sampling fully explores parameter space and can capture non-Gaussian uncertainties and non-linear correlations between parameters without assuming they are independent, and accounts for covariances and any prior knowledge.

Efficient MCMC requires the stellar models to be run in seconds or less, but main-sequence simulations take minutes to hours to produce converged models with suitable age resolution.~To make the Bayesian inference computationally feasible, we train a machine learning (ML) surrogate for stellar structure code models as in \citet{Dennis2025a, Dennis2025b, Franz2023} and similar to \citet{Yang2026, Grichener2025, Vermarien2025, Tahseen2025, Scoggins2025, Conceicao2024, Doeser2024, SpurioMancini2022} and other works which have created and applied ML surrogates to other areas of astronomy (see \citet{Ting2025}'s section 3.2 for a review on simulation based inference with deep learning). To improve the efficiency of training grid creation, we use an active learning (AL) strategy that selectively evaluates new models where the surrogate uncertainty is largest. Applying our method to 38 stars in MS DEB systems, we obtain inferences of their initial mass ($M$), initial metallicity ($Z$), convective mixing length parameter ($\alpha_{\rm MLT}$), and stellar age ($\tau$).

\par This paper is organized as follows. We discuss MS DEBs and the data we use in \S \ref{sec:observational_data}, our technique for emulating the main sequence in \S \ref{sec:emulating}, the details of our MCMC analysis in \S \ref{sec:mcmc}, and present our results in \S \ref{sec:results}. We discuss our results in \S \ref{sec:discussion} and draw conclusions in \S \ref{sec:conclusion}.

\section{Methodology}
\label{sec:methodology}
\par This section is split into the following subsections: \S \ref{sec:observational_data} contains the necessary empirical data and results that we used for this work; \S \ref{sec:emulating} describes the process of emulating the stellar structure code including fixed input physics and the specifics of selecting new models; and \S \ref{sec:evaluating} presents our final evaluation of our ML surrogate.

\subsection{Observational Data Selection}
\label{sec:observational_data}
\par
\begin{figure}
    \centering
    \includegraphics[width=0.42\textwidth]{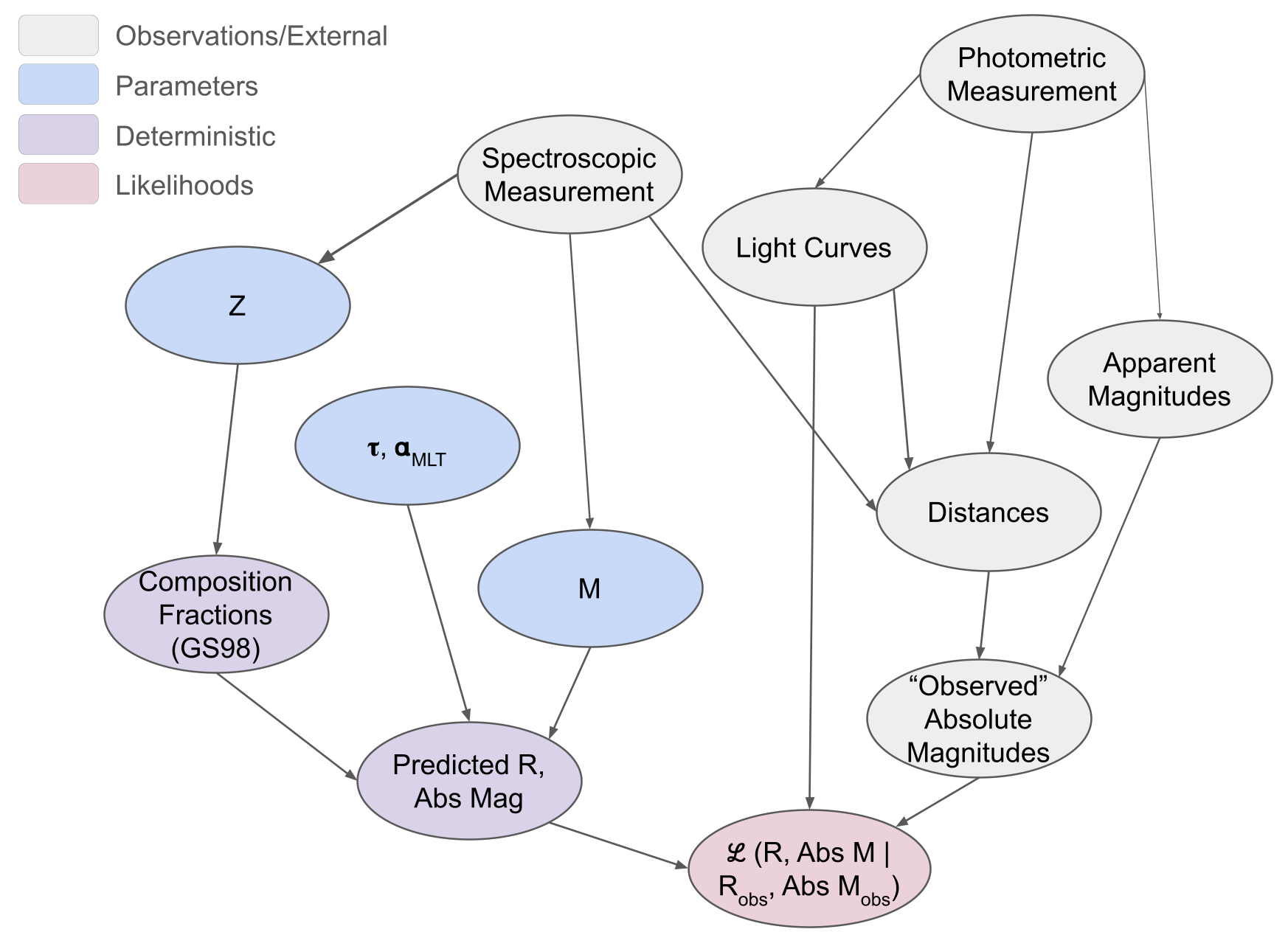}
    \caption{A directed acyclic graph showing the connection between the parameters and the observables. This graph forms the probabilistic model used to forward model the MS DEB magnitudes and radii.}
    \label{fig:dag}
\end{figure}

 The data we need to infer are $M$, $Z$, $\alpha_{\rm MLT}$ and $\tau$ with a Bayesian model from flux measurements in the $M_B$ and $M_V$ band passes and spectroscopic measurements from each system. From these sets of measurements of binary systems we can obtain Doppler shifts to calculate orbital velocities and orbital periods to infer individual masses, utilize radial velocities and orbital separation to infer individual radii, and use elemental emission line ratios to infer metallicity fractions. Because of the unique precision of binary systems, these $M$, $Z$, and $R$ values provide extra information in the priors and likelihood of our MCMC. A table of our free parameters and the priors we utilize for each are provided in Table \ref{tbl:parameters}. A directed acyclic graph depicting how these parameters and observables are connected is provided as Figure \ref{fig:dag}.

\begin{table}
    \centering
    \begin{tabular}{|c|c|c|}
        Free Parameter & Prior & Description \\
        \hline
        $M$ & $\mathcal{N}(\mu_{M,i}, \sigma_{M,i})$ & Initial Mass \\
        \hline
        $Z$ & $\mathcal{N}(\mu_{Z,i}, \sigma_{Z,i})$ & Initial Metallicity \\
        \hline
        $\alpha_{\rm MLT}$ & $\mathcal{U}(0.01, 3.0)$ & Mixing Length Parameter \\
        \hline
        $\tau$ & $\mathcal{U}(0, \sim 14 \times 10^9) $ & Stellar Age \\
        \hline
    \end{tabular}
    \caption{Inferred parameters in our study. For $M$ and $Z$, $i$ indexes the star (a row in Table \ref{tbl:measurements}). The values reported for $M$ and $Z$ in Table \ref{tbl:measurements} are $\mu$ and their errors are $\sigma$ used in the $\mathcal{N}(\mu,\sigma)$ function.}
    \label{tbl:parameters}
\end{table}

\par
\begin{figure*}
    \centering
    \includegraphics[width=\linewidth]{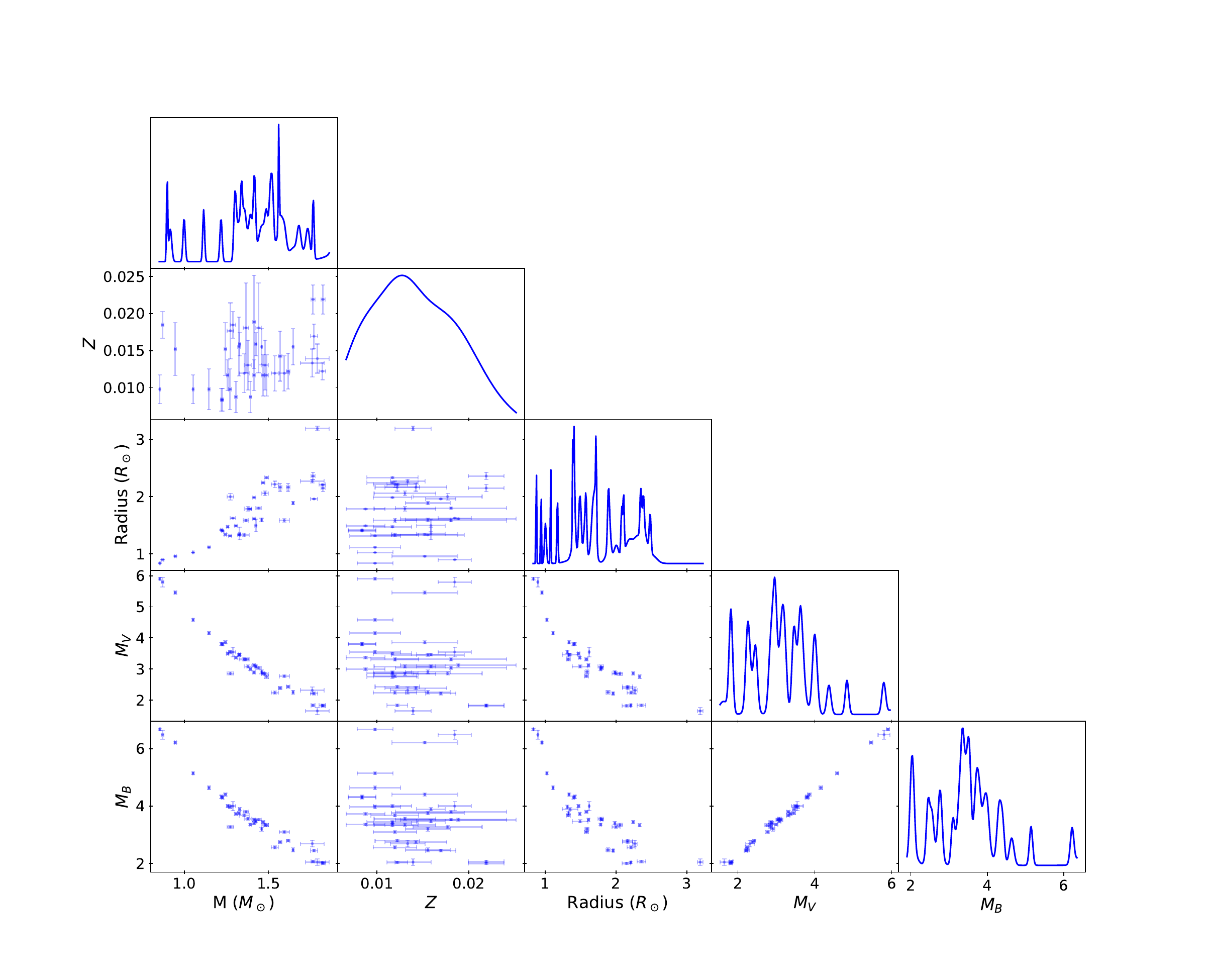}
    \caption{Corner plot of the stellar data used in this study (Table \ref{tbl:measurements}) with scatter plots on the off-diagonal and kernel density estimates of the 1D distributions along the diagonal. The latter adopt a bandwidth equal to the uncertainty in each case.The error bars show 68\% credible intervals (which may be terminated by a prior boundary), and intersect at the median posterior values.}
    \label{fig:data_distribution}
\end{figure*}

 We utilize $M$, $Z$, and $R$ data from \cite{Eker2018}\footnote{\label{ftn:vizier} Downloaded from \href{Vizier}{https://vizier.cds.unistra.fr/viz-bin/VizieR?-source=J/MNRAS/479/5491}}, and photometric $M_V$ and $M_B$ data from \cite{Bakis2022}. These data are collated in Table \ref{tbl:measurements} and the corner plot in Figure \ref{fig:data_distribution} summarizes this data. This table also includes $\log(g)$ and $T_{\rm eff}$ values that we refer to later in \S \ref{sec:discussion}. We use the 38 stars from Table \ref{tbl:measurements}, specifically their estimates in $M$ and $Z$, the estimate of their radii, and their measurements in $M_V$, and $M_B$ directly in our model as shown in Figure \ref{fig:dag}. 

\begin{table*}
    %\centering
    \begin{tabular}{|c|c|c|c|c|c|c|c|}
        \hline 
Star & $M$ ($M_\odot$) & $Z$ & $R$ ($R_\odot$) & $\log(g)$ & $T_\textrm{eff}$ & $M_V$ & $M_B$ \\
\hline
V421 Peg p & $1.594 \pm 0.029$ & $0.012 \pm 0.002$ & $1.584 \pm 0.028$ & $4.241 \pm 0.017$ & $7250 \pm 80$ & $2.770 \pm 0.042$ & $3.096 \pm 0.042$\\ 
\hline
V421 Peg s & $1.356 \pm 0.029$ & $0.012 \pm 0.003$ & $1.328 \pm 0.029$ & $4.324 \pm 0.021$ & $6980 \pm 120$ & $3.310 \pm 0.042$ & $3.673 \pm 0.042$\\ 
\hline
YZ Cas s & $1.325 \pm 0.007$ & $0.016 \pm 0.004$ & $1.331 \pm 0.006$ & $4.312 \pm 0.005$ & $6880 \pm 240$ & $3.453 \pm 0.044$ & $3.754 \pm 0.049$\\ 
\hline
V570 Per p & $1.425 \pm 0.006$ & $0.016 \pm 0.002$ & $1.494 \pm 0.110$ & $4.243 \pm 0.064$ & $6842 \pm 25$ & $3.079 \pm 0.045$ & $3.466 \pm 0.051$\\ 
\hline
V570 Per s & $1.328 \pm 0.006$ & $0.016 \pm 0.002$ & $1.354 \pm 0.110$ & $4.298 \pm 0.071$ & $6580 \pm 70$ & $3.463 \pm 0.045$ & $3.890 \pm 0.051$\\ 
\hline
V1130 Tau p & $1.306 \pm 0.008$ & $0.009 \pm 0.002$ & $1.490 \pm 0.010$ & $4.208 \pm 0.006$ & $6650 \pm 70$ & $3.368 \pm 0.041$ & $3.726 \pm 0.041$\\ 
\hline
V1130 Tau s & $1.392 \pm 0.008$ & $0.009 \pm 0.002$ & $1.784 \pm 0.011$ & $4.079 \pm 0.006$ & $6625 \pm 70$ & $2.994 \pm 0.041$ & $3.355 \pm 0.041$\\ 
\hline
CD Tau p & $1.441 \pm 0.016$ & $0.018 \pm 0.006$ & $1.798 \pm 0.017$ & $4.087 \pm 0.010$ & $6200 \pm 50$ & $3.036 \pm 0.042$ & $3.526 \pm 0.042$\\ 
\hline
CD Tau s & $1.366 \pm 0.016$ & $0.018 \pm 0.006$ & $1.584 \pm 0.020$ & $4.174 \pm 0.012$ & $6200 \pm 50$ & $3.311 \pm 0.042$ & $3.801 \pm 0.042$\\ 
\hline
V501 Mon p & $1.645 \pm 0.004$ & $0.016 \pm 0.002$ & $1.888 \pm 0.029$ & $4.103 \pm 0.013$ & $7510 \pm 100$ & $2.252 \pm 0.059$ & $2.472 \pm 0.066$\\ 
\hline
V501 Mon s & $1.459 \pm 0.003$ & $0.016 \pm 0.002$ & $1.592 \pm 0.028$ & $4.198 \pm 0.015$ & $7000 \pm 90$ & $2.907 \pm 0.059$ & $3.195 \pm 0.066$\\ 
\hline
GX Gem p & $1.488 \pm 0.011$ & $0.012 \pm 0.003$ & $2.334 \pm 0.012$ & $3.875 \pm 0.006$ & $6194 \pm 100$ & $2.752 \pm 0.053$ & $3.331 \pm 0.054$\\ 
\hline
GX Gem s & $1.467 \pm 0.010$ & $0.012 \pm 0.003$ & $2.244 \pm 0.012$ & $3.903 \pm 0.006$ & $6166 \pm 100$ & $2.858 \pm 0.053$ & $3.443 \pm 0.054$\\ 
\hline
VZ Hya p & $1.271 \pm 0.009$ & $0.010 \pm 0.003$ & $1.315 \pm 0.005$ & $4.305 \pm 0.005$ & $6645 \pm 150$ & $3.546 \pm 0.049$ & $3.969 \pm 0.060$\\ 
\hline
VZ Hya s & $1.146 \pm 0.006$ & $0.010 \pm 0.003$ & $1.113 \pm 0.007$ & $4.405 \pm 0.006$ & $6290 \pm 150$ & $4.155 \pm 0.049$ & $4.637 \pm 0.060$\\ 
\hline
RS Cha p & $1.823 \pm 0.012$ & $0.022 \pm 0.002$ & $2.150 \pm 0.060$ & $4.034 \pm 0.024$ & $7638 \pm 76$ & $1.814 \pm 0.041$ & $1.999 \pm 0.041$\\ 
\hline
RS Cha s & $1.764 \pm 0.012$ & $0.022 \pm 0.002$ & $2.360 \pm 0.060$ & $3.939 \pm 0.022$ & $7228 \pm 72$ & $1.830 \pm 0.041$ & $2.067 \pm 0.041$\\ 
\hline
ZZ Boo p & $1.616 \pm 0.010$ & $0.012 \pm 0.002$ & $2.164 \pm 0.070$ & $3.976 \pm 0.028$ & $6860 \pm 20$ & $2.429 \pm 0.042$ & $2.794 \pm 0.042$\\ 
\hline
ZZ Boo s & $1.568 \pm 0.010$ & $0.014 \pm 0.003$ & $2.164 \pm 0.070$ & $3.963 \pm 0.028$ & $6930 \pm 20$ & $2.386 \pm 0.042$ & $2.741 \pm 0.042$\\ 
\hline
V636 Cen p & $1.052 \pm 0.005$ & $0.010 \pm 0.002$ & $1.024 \pm 0.004$ & $4.440 \pm 0.004$ & $5900 \pm 85$ & $4.583 \pm 0.042$ & $5.143 \pm 0.042$\\ 
\hline
V636 Cen s & $0.854 \pm 0.003$ & $0.010 \pm 0.002$ & $0.835 \pm 0.004$ & $4.526 \pm 0.004$ & $5000 \pm 100$ & $5.901 \pm 0.042$ & $6.677 \pm 0.042$\\ 
\hline
AD Boo p & $1.414 \pm 0.009$ & $0.019 \pm 0.006$ & $1.614 \pm 0.014$ & $4.173 \pm 0.008$ & $6575 \pm 120$ & $3.120 \pm 0.050$ & $3.523 \pm 0.052$\\ 
\hline
WZ Oph p & $1.227 \pm 0.007$ & $0.008 \pm 0.002$ & $1.401 \pm 0.012$ & $4.234 \pm 0.008$ & $6165 \pm 100$ & $3.796 \pm 0.047$ & $4.300 \pm 0.056$\\ 
\hline
WZ Oph s & $1.220 \pm 0.006$ & $0.008 \pm 0.002$ & $1.419 \pm 0.012$ & $4.221 \pm 0.008$ & $6115 \pm 100$ & $3.807 \pm 0.047$ & $4.320 \pm 0.056$\\ 
\hline
V2653 Oph p & $1.537 \pm 0.021$ & $0.012 \pm 0.002$ & $2.215 \pm 0.055$ & $3.934 \pm 0.022$ & $6950 \pm 480$ & $2.237 \pm 0.047$ & $2.557 \pm 0.047$\\ 
\hline
V2653 Oph s & $1.273 \pm 0.019$ & $0.018 \pm 0.004$ & $2.000 \pm 0.056$ & $3.941 \pm 0.025$ & $6350 \pm 650$ & $2.855 \pm 0.047$ & $3.270 \pm 0.047$\\ 
\hline
KIC 9851944 p & $1.760 \pm 0.070$ & $0.013 \pm 0.002$ & $2.270 \pm 0.030$ & $3.972 \pm 0.021$ & $7026 \pm 100$ & $2.312 \pm 0.111$ & $2.689 \pm 0.111$\\ 
\hline
KIC 9851944 s & $1.790 \pm 0.070$ & $0.014 \pm 0.002$ & $3.190 \pm 0.040$ & $3.684 \pm 0.020$ & $6902 \pm 100$ & $1.648 \pm 0.111$ & $2.043 \pm 0.111$\\ 
\hline
OO Peg p & $1.820 \pm 0.020$ & $0.012 \pm 0.001$ & $2.210 \pm 0.010$ & $4.010 \pm 0.006$ & $7850 \pm 350$ & $1.827 \pm 0.043$ & $2.035 \pm 0.042$\\ 
\hline
OO Peg s & $1.770 \pm 0.020$ & $0.017 \pm 0.002$ & $1.960 \pm 0.010$ & $4.102 \pm 0.007$ & $7600 \pm 450$ & $2.212 \pm 0.043$ & $2.449 \pm 0.042$\\ 
\hline
V375 Cep p & $1.288 \pm 0.017$ & $0.018 \pm 0.002$ & $1.623 \pm 0.006$ & $4.128 \pm 0.007$ & $6230 \pm 50$ & $3.546 \pm 0.154$ & $3.998 \pm 0.154$\\ 
\hline
V375 Cep s & $0.871 \pm 0.008$ & $0.018 \pm 0.002$ & $0.897 \pm 0.003$ & $4.473 \pm 0.005$ & $5151 \pm 50$ & $5.800 \pm 0.154$ & $6.490 \pm 0.154$\\ 
\hline
BW Aqr p & $1.479 \pm 0.020$ & $0.013 \pm 0.003$ & $2.057 \pm 0.040$ & $3.982 \pm 0.018$ & $6350 \pm 100$ & $2.845 \pm 0.049$ & $3.342 \pm 0.053$\\ 
\hline
BW Aqr s & $1.377 \pm 0.020$ & $0.013 \pm 0.003$ & $1.788 \pm 0.040$ & $4.073 \pm 0.020$ & $6450 \pm 100$ & $3.078 \pm 0.049$ & $3.558 \pm 0.053$\\ 
\hline
EF Aqr p & $1.244 \pm 0.008$ & $0.015 \pm 0.004$ & $1.338 \pm 0.012$ & $4.280 \pm 0.008$ & $6150 \pm 65$ & $3.858 \pm 0.046$ & $4.403 \pm 0.046$\\ 
\hline
EF Aqr s & $0.946 \pm 0.006$ & $0.015 \pm 0.004$ & $0.956 \pm 0.012$ & $4.453 \pm 0.011$ & $5185 \pm 110$ & $5.458 \pm 0.046$ & $6.217 \pm 0.046$\\ 
\hline
BK Peg p & $1.414 \pm 0.007$ & $0.012 \pm 0.002$ & $1.985 \pm 0.008$ & $3.993 \pm 0.004$ & $6265 \pm 85$ & $2.888 \pm 0.043$ & $3.400 \pm 0.058$\\ 
\hline
BK Peg s & $1.257 \pm 0.005$ & $0.012 \pm 0.002$ & $1.472 \pm 0.017$ & $4.202 \pm 0.010$ & $6320 \pm 30$ & $3.497 \pm 0.043$ & $3.999 \pm 0.058$\\ 
\hline
    \end{tabular}
    \caption{A list of the {inputs to our analysis}. $M$ and $Z$ are used as priors in the MCMC and $R$, $M_V$, and $M_B$ are treated as observables in the likelihood. The values for $M$, $R$, $\log(g)$ and $T_{\rm eff}$ were pulled from Table 1 of \protect\cite{Eker2018} and the $Z$ from their Table 2 downloaded via \textit{Vizier}\protect\footref{ftn:vizier}, and the $M_V$ and $M_B$ values are from Table 3 of \protect\cite{Bakis2022}.}
    \label{tbl:measurements}
\end{table*}

\subsection{Emulating the Main Sequence}
\label{sec:emulating}
As noted above, to enable efficient exploration of parameter space with MCMC, we replace direct stellar evolution calculations with a machine learning emulator. The emulator predicts stellar radius ($R$) and magnitudes $M_V$ and $M_B$ (which are produced by the bolometric corrections built into the stellar structure code) as functions of the stellar parameters $M$, $Z$, $\alpha_{\rm MLT}$, and {$\tau$}. The emulator is trained on a set of stellar evolution models generated with the stellar structure code Modules for Experiments in Stellar Astrophysics (MESA) version 12778 \citep{MESA2011,MESA2013,MESA2015,MESA2018,MESA2019,MESA2023}. In the following subsections we describe the construction of the training set and the active learning strategy used to efficiently sample the relevant regions of parameter space.

\begin{figure*}
    \centering
    \includegraphics[width=0.9\linewidth]{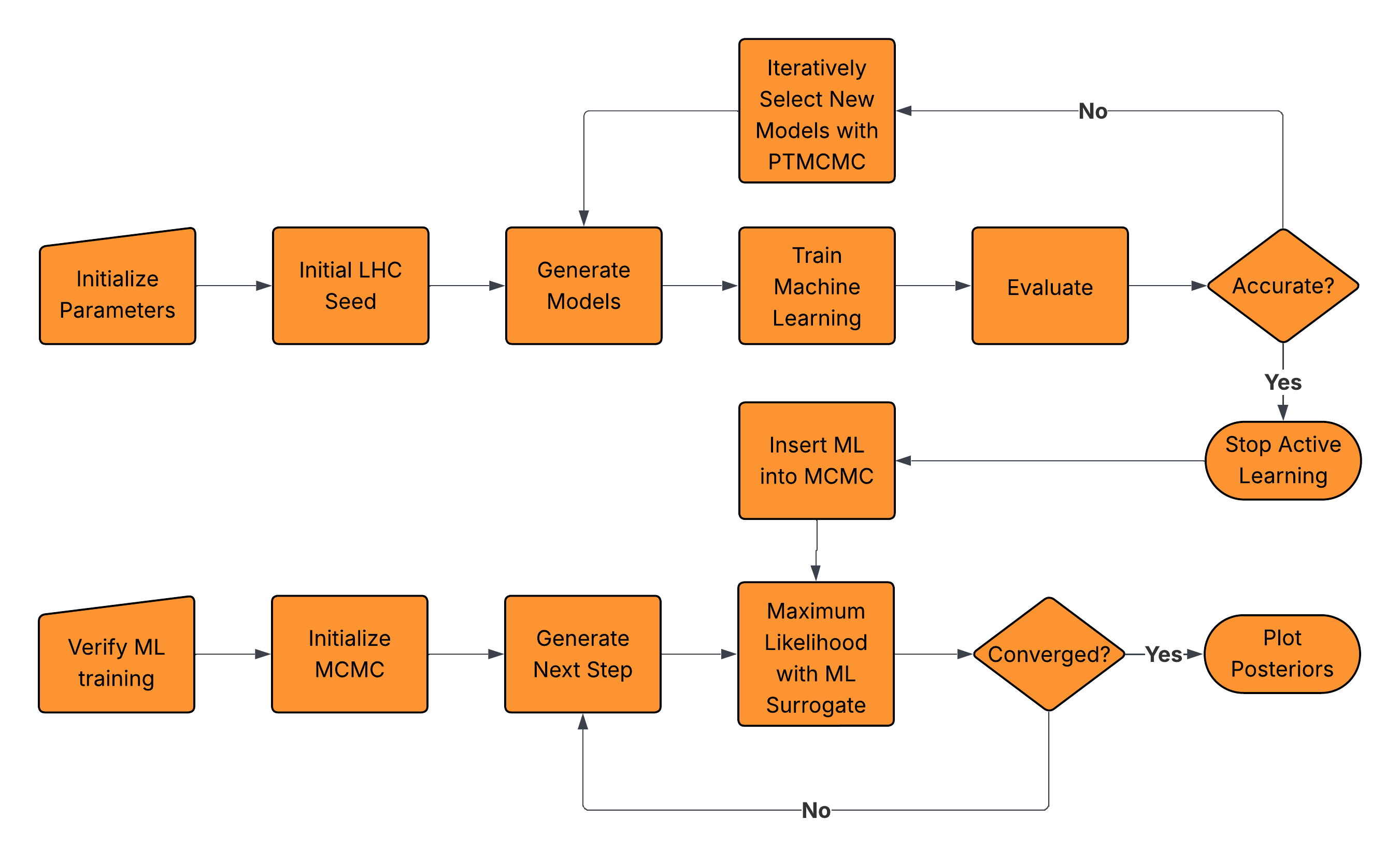}
    \caption{A flowchart depicting the entire active learning procedure. The upper half of the diagram outlines the data selection and generation procedure used to curate the set of models which trained the ML surrogate. The lower half outlines the procedure used once the ML training was completed and the final MCMC --- the one used to constrain parameters --- is run to generate posteriors across the input parameters. The procedure begins with a Latin-Hyper-Cube-sampled set of models and then iteratively trains using a parallel tempering markov chain monte carlo (PTMCMC) sampling simulation to select new training data until a threshold accuracy is reached. }
    \label{fig:flow_diagram}
\end{figure*}

\par Information on our choice of input physics and the stellar simulator we used can be found in \S \ref{sec:input_physics}, the procedure for selecting new training models are in \S \ref{sec:generating}, and our evaluation of the machine learning surrogates' performance can be found in \S \ref{sec:evaluating}. 

\subsubsection{Input Physics}
\label{sec:input_physics}
We evolve our models from the pre-main-sequence through the main sequence, and terminate them at the turn off of core hydrogen burning, which we defined as the point where the total central pp chain burning had decayed to less than 50\% of its maximum. We vary the input parameters $M$, $Z$, and $\alpha_{\rm MLT}$ and hold all other parameters fixed. The maximum timestep is fixed to $10^7$ years after the star reaches the zero age main sequence (ZAMS), defined as the time when the difference between the surface and hydrogen burning luminosity becomes less than 0.0001. This forces MESA to output at roughly $10^7$ year intervals contributing many models for a single set of input parameters.

Our models adopt the following input physics:
\begin{description}
    \item \textbf{Abundances:} GS98 Solar composition \citep{GS98}.
    \item \textbf{Opacities:} MESA default OPAL opacities \citep{Iglesias1996, Iglesias1993} with type 2 opacities from \citep{Cassisi2007}.
    \item \textbf{Nuclear reaction rates:} JINA REACLIB \citep{Cyburt2010}.
    \item \textbf{Equation of state:} OPAL 2005 \citep{Rogers2002}.
    \item \textbf{Convective mixing:} Cox mixing length formulation \citep{Cox1968}.
\end{description}
A template of MESA inlists used to generate the models is provided in the accompanying reproduction package.

\subsubsection{Grid Generation}
\label{sec:generating}
As noted above, to maximize computational efficiency we adapt an active learning scheme to choose which models to evaluate. We use the following prescription, similar to \cite{Rocha2022}, as our active learning procedure.
\begin{enumerate}[left=0pt..20pt]
    \item Initial seeding
    \item Train ML on models
    \item Generate new samples to improve error
    \begin{enumerate}[itemindent=!]
        \item Repeated sampling with MCMC on an acquisition function
        \item First sample on classification task, rerun MESA, retrain ML and repeat until the error threshold is met.
        \item Next sample on regression task, rerun MESA, retrain ML and repeat until the error threshold is met.
    \end{enumerate}
\end{enumerate}
The first phase is a Latin Hyper-Cube (LHC) sample \citep{McKay1979} initialization to select the first 1000 models. We draw LHC samples in mass from $0.6-2.2 \msun$, in Z from $0.0001-0.04$, and $\alpha_{\rm MLT}$ from $0.01 - 3.0$.~We remind the reader that the ages are recorded in time-steps of no greater than $10^7$ years post-ZAMS.

Using these initial seed models, we train a neural network classifier and a neural network regressor for our active learning procedure. For all of our ML algorithms, we divide our MESA data into training, validation, and testing subsets with an 80\%/10\%/10\% split.

\par We choose neural networks for the AL procedure because they provide a number of advantages over other ML techniques. Primarily, to successfully perform AL, we must be able to provide some reasonable estimate of the error in the ML algorithm on unlabeled data (i.e., input parameters for which we do not have MESA model outputs). For classification tasks this is fairly straightforward (see Eq.~\eqref{eq:classifier_acquisition}), but for regression providing an estimate of the error is more complicated. While the typical approach would be to fit a Gaussian process (GP) instead of a neural network, GPs do not scale well to large datasets and \textit{a priori} we have no guarantee our dataset will not reach {a size prohibitive to their use}. Conversely, neural networks do scale well to large data and can be approximated to GPs, allowing us to estimate their errors \citep{Gal2015}.

All our machine learning models take $M$, $Z$, $\alpha_{\rm MLT}$, and {$\log_{10}(\tau)$} as inputs. Because the inference explores the full parameter space, some proposed combinations correspond to stellar models that are not on the main-sequence. To avoid evaluating the regression emulator outside its domain of validity, we first train a classification model that identifies whether a given set of input parameters corresponds to a main-sequence star. We incorporate this classifier into the active learning procedure to ensure that training samples and MCMC proposals remain within the main-sequence region.

\par To select which sets of new input parameters to simulate, we use a scheme that attempts to maximize the improvement of the network with each additional MESA model run. We do this by using a MCMC algorithm that uses uniform priors across all parameters and substitutes the likelihood function of the MCMC with an acquisition function designed to find the parameters where the neural networks produce the most error or have the least confidence. Because we expect the posteriors for the acquisition functions to be multi-modal, we use parallel tempering Markov-Chain Monte Carlo (PTMCMC) throughout our active learning procedure, using a similar implementation to \citet{Rocha2022}.

\par The acquisition function we use for the classification task is identical to the one used by \cite{Rocha2022}, defined as
\begin{equation}
    \label{eq:classifier_acquisition}
    A(\theta) = 1 - \textrm{max}(\textrm{ML}(\theta)),
\end{equation}
where $\theta$ is a set of $M$, $Z$, $\alpha_{\rm MLT}$, and {$\log_{10}(\tau)$}, and $\textrm{ML}(\theta)$ is the output of the neural network classifier's softmax output layer, i.e. the probability of the input data belonging to a specific class. 

\begin{figure}
    \centering
    \includegraphics[scale=0.5]{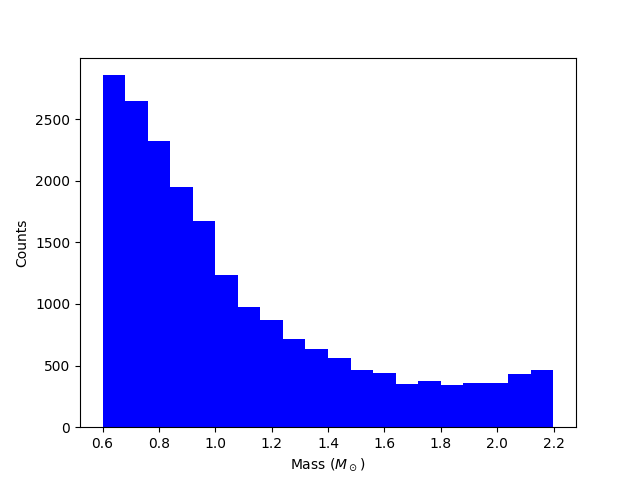}
    \includegraphics[scale=0.5]{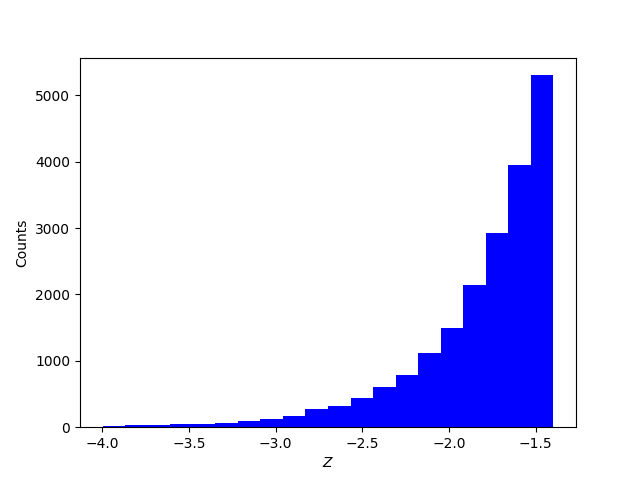}
    \caption{Histograms showing the distributions across individual parameters space at the completion of the AL procedure. The neural network emulator is only  valid across parts of parameter space represented in these plots. They also show where the AL procedure determined where the most models were needed (i.e. at the lower end of our $M$ range and the upper end of our $Z$ range). Not shown is the distribution in $\alpha_{\rm MLT}$, which is uniform.}
    \label{fig:parameter_space}
\end{figure}

By using Equation \eqref{eq:classifier_acquisition} as our likelihood in the MCMC function, the walkers of the AL stage MCMC prefer areas of parameter space where the surrogate model's classifier is most uncertain about its classification. We expect this to yield a multi-modal solution with the centers of the modes near the boundaries between the three classes (on the main sequence, not on the main sequence, too old for the age of the universe). Using this function, we avoid regions of parameter space where the classifier is highly confident and does not need more training data in favor of areas of greater uncertainty. One notable drawback of this method is that the neural network surrogate may still be making large errors, but with high confidence so the acquisition function does not explore that part of parameter space. However, we do not expect this to impact our results because the priors we apply later in \S \ref{sec:mcmc} are very restrictive, and as we show in Figure \ref{fig:parameter_space}, we fully explore the relevant parameter space within our training data.

\par Once the surrogate model's classifier has reached a sufficiently high accuracy (97\%), we then switch priorities to the regressor. We remind the reader that even if new MESA model selection prioritizes either the surrogate regressor or classifier, the data from the evaluated models can be used to train both.  We also note that because the surrogate's regressor only needs to be accurate on models that are on the MS, we apply the classification of the surrogate model's classifier as a prior for the MCMC used at this stage.

\par We again use $M$, $Z$, $\alpha_{\rm MLT}$, and {$\log_{10}(\tau)$} as inputs and output $M_V, M_B$, and $R$. The acquisition function we used to select new MESA models for the regressor is the expected improvement (EI) from Bayesian optimization and is calculated with,
\begin{multline}
    \label{eq:regressor_acquisition}
    A(\theta) = \Sigma_{i}(\mu_i - \theta_i - \xi_{EI}) \times \frac{1}{2} \left(1 + \textrm{erf}\left(\frac{\left(\frac{\mu_i - \theta_i - \xi_{EI}}{\sigma_i}\right)}{\sqrt{2}}\right)\right) + \\
    \sigma_i \times \mathcal{N}\left(0, \left(\frac{(\mu_i - \theta_i - \xi_{EI})}{\sigma_i}\right)\right),
\end{multline}
which is the equation for the EI of a GP (see \cite{Leclercq2018} and \cite{Brochu2010} for the derivation) summed over each output $i$ ($M_V, M_B$, and $R$) for some predicted value of $\theta$ which was evaluated over a single set of parameters. In this equation, $\mu_i$ is the average of our deep neural network outputs over repeated use of the same inputs and $\sigma_i$ is the standard deviation of those outputs, and $\xi_{EI}$ is a cooling factor described by \cite{Brochu2010} which we set to zero throughout our procedure. To acquire the mean $\mu_i$ and standard deviation $\sigma_i$ used in \eqref{eq:regressor_acquisition}, we evaluate the neural network over the same parameters which output $\theta$, but instead of evaluating the neural network in inference mode (which was used for $\theta$), we evaluate it in training mode. Because random dropout \citep{Srivastava2014} layers were used to construct the neural network, when evaluated in training mode many times (in this case 100), the neural network will produce 100 different outputs whose mean and standard deviation can be approximated to the mean and standard deviation of a gaussian process \citep{Gal2015}. By using this function, it is possible to trade off between exploration and exploitation by adjusting the $\xi_{EI}$ parameter and hence optimize our selection of new MESA models. By default we set  $\xi_{EI}=0.0$.~{While we do not vary it, $\xi$ can be adjusted at each iteration of the active learning procedure if desired}. During this step, we also parameterize the error of the machine learning by training a random forest to output the residuals of the neural network regressor (i.e., the difference between the true MESA output values and its predictions). By using the MCMC in this way, we determine which regions of parameter space are more likely to yield larger improvements in our machine learning surrogate. 

\par When the MCMC is finished for either the classification step or the regression step, we discard burn-in and take a subsample of the posteriors which have been produced with either \eqref{eq:classifier_acquisition} or \eqref{eq:regressor_acquisition}. This subsample is then passed to MESA and the results added to a master dataset of MESA simulation results. However, for the regressor, the parameters of this subsample are first passed to the random forests that we trained to parameterize its error and plotted as corner plots. This allows us to visually inspect the error in the active learning regressor in the areas where the acquisition function is maximizing for improvement, i.e., in the places we expect the error to be the highest across parameter space. We can use this to evaluate the expected regression error using the parameterized error models. If the error is subdominant to our observational errors (e.g. an order of magnitude smaller than the average observational error.), we terminate the AL procedure, and if not we pull samples from the chain to run in MESA.

\par This procedure is iterated until first the desired classification and then the desired regression accuracy are reached.

\subsection{Training and Evaluating the Final ML Surrogate}
\label{sec:evaluating}

\par While it was necessary to use neural networks for the AL procedure, we used a random forest for our final MESA emulator because these tend to perform better on tabular format data and have been shown to be more accurate than neural networks on MESA data \citep{Dennis2025a}.

\par To quantify the uncertainty of the machine learning surrogate, we trained an additional random forest to model the residuals between the emulator predictions and the MESA outputs. The residual model was trained on the validation dataset that was held out from the original training set, and its performance was evaluated on the testing dataset.~Because machine-learning models typically perform better when trained to predict a single target variable, we trained separate emulators for each stellar observable, $R$, $M_V$, and $M_B$, and their residual errors. This amounts to six random forests in total.

\par Our final parameter list comprises 20,000 sets of $M$, $Z$, and $\alpha_{\rm MLT}$. Of these, a small number failed to converge or did not reach the end of the MS within 12 hours. We implemented this 12 hour cutoff to maintain computational efficiency as the average runtime for models is of order minutes or less. Our final dataset is over one million data points when the evolutionary time-steps are included. The random forest classifier achieves an accuracy of 99.8\%. The regression random forests have root-mean-square errors (RMSE) of 0.00257 for the radius, 0.00263 for the $V$-band magnitude, and 0.00297 for the $B$-band magnitude. These errors are quoted in normalized units, since the models were trained on data scaled to the interval [0,1]. When converted to physical units, the corresponding residuals can be directly compared with the observational uncertainties. 

\par For each surrogate regressor we compare the distribution of its residuals with the observational uncertainties. The residuals of the machine learning surrogate are significantly smaller than the observational errors for all three observables ($R$, $M_V$, and $M_B$), indicating that the emulator error is subdominant.~Specifically, the 16th–84th percentile range of the surrogate residuals is $\lesssim 10^{-3}$ for all observables, whereas the observational uncertainties are typically $\sim$$10^{-2}$–$10^{-1}$.~We note that these residuals are evaluated over the full parameter space spanned by the training set, whereas the final MCMC chains will explore only a restricted region of this space. The machine learning uncertainty is therefore conservatively estimated and is explicitly included in the likelihood (see \S\ref{sec:likelihood}).

\subsection{Constraining Parameters}
\label{sec:mcmc}
\par In this section we attempt to constrain the values of $\alpha_{\rm MLT}$ and {$\tau$} for each of the stars in our study. We do this by evaluating MCMC using our trained random forest surrogate described in \S \ref{sec:evaluating} in place of MESA. We use priors on our parameters from the $M$ and $Z$ columns in Table \ref{tbl:measurements} (described in \S \ref{sec:priors}) and use the $R$, $M_V$, and $M_B$ from Table \ref{tbl:measurements} in our likelihood (described in \S \ref{sec:likelihood}).

\subsubsection{Priors}
\label{sec:priors}
\par Our priors are as follows. For $M$ and $Z$ we apply gaussian priors with mean and standard deviation which are reported as the value and errors in Table \ref{tbl:measurements}. We use uniform priors for $\alpha_{\rm MLT}$ and {$\tau$} since no prior knowledge about their values are available.\footnote{Although previous studies have inferred $\alpha_{\rm MLT}$ for particular stellar populations (e.g.\ \citealt{Magic2015, Song2020, Desmond2021}), we adopt a uniform prior so that the DEB data alone determine the constraints.} Additionally, we treat the random forest surrogate's classifier as a prior. The classifier, which distinguishes between pre-MS models, MS models, and post-MS models assigns a prior of zero for models that do not fall on the MS. This is a necessary step because the MCMC may choose parameters that do not lie on the main sequence, and has the added benefit of helping ensure the MCMC does not stray into regions of parameter space upon which our regression algorithm was not trained.

\subsubsection{Likelihood}
\label{sec:likelihood}

In the inference stage, the trained surrogate model replaces MESA when evaluating the likelihood. For a given set of parameters $\theta$, the emulator predicts the observables ($R$, $M_V$, $M_B$), which are compared with the measured values. We adopt the following Gaussian likelihood,
\begin{equation}
    \label{eq:likelihood}
    -2\ln(\mathcal{L}) = \sum_{\textrm{i=outputs}} \left(\frac{(\textrm{ML}(\theta)_i - y_i)^2}{(\sigma_{y,i}^2 + \sigma_{\textrm{ML}, i}^2)}\right) + \ln(2\pi (\sigma_{y,i}^2 + \sigma_{\textrm{ML}, i}^2)),
\end{equation}
where $\textrm{ML}(\theta)_i$ is one of the output values $i$ of the ML regressor for a particular set of inputs (i.e., $i$ is $R$ or one of our magnitudes), $y_i$ is the observed value for output column $i$, $\sigma_{y, i}$ is the observational error for output column $i$, and $\sigma_{\textrm{ML}, i}$ is the machine learning error for output $i$. To determine the value for $\sigma_{\textrm{ML}, i}$ we fed the inputs to the random forest parameterization of the error from \S \ref{sec:evaluating} above. We utilized the python MCMC package emcee \cite{ForemanMackey2013} and for each model we ran 32 walkers simultaneously, calculating the autocorrelation time between the walkers every 5,000 samples.

\par We determined the chains had converged when the average integrated autocorrelation time (i.e., the number of steps a chain in the MCMC needs to ``forget" where it started) was less than 0.1\% of the total chain length and had changed less than 1\% over the previous 5,000 samples \cite{ForemanMackey2013}. Once the convergence condition was met, we terminated the MCMC. We discard twice the maximum value of the final autocorellation time as burn-in \citet{ForemanMackey2013}.

\section{Results}
\label{sec:results}
\par
\begin{figure}
    \centering
    \includegraphics[scale=0.35]{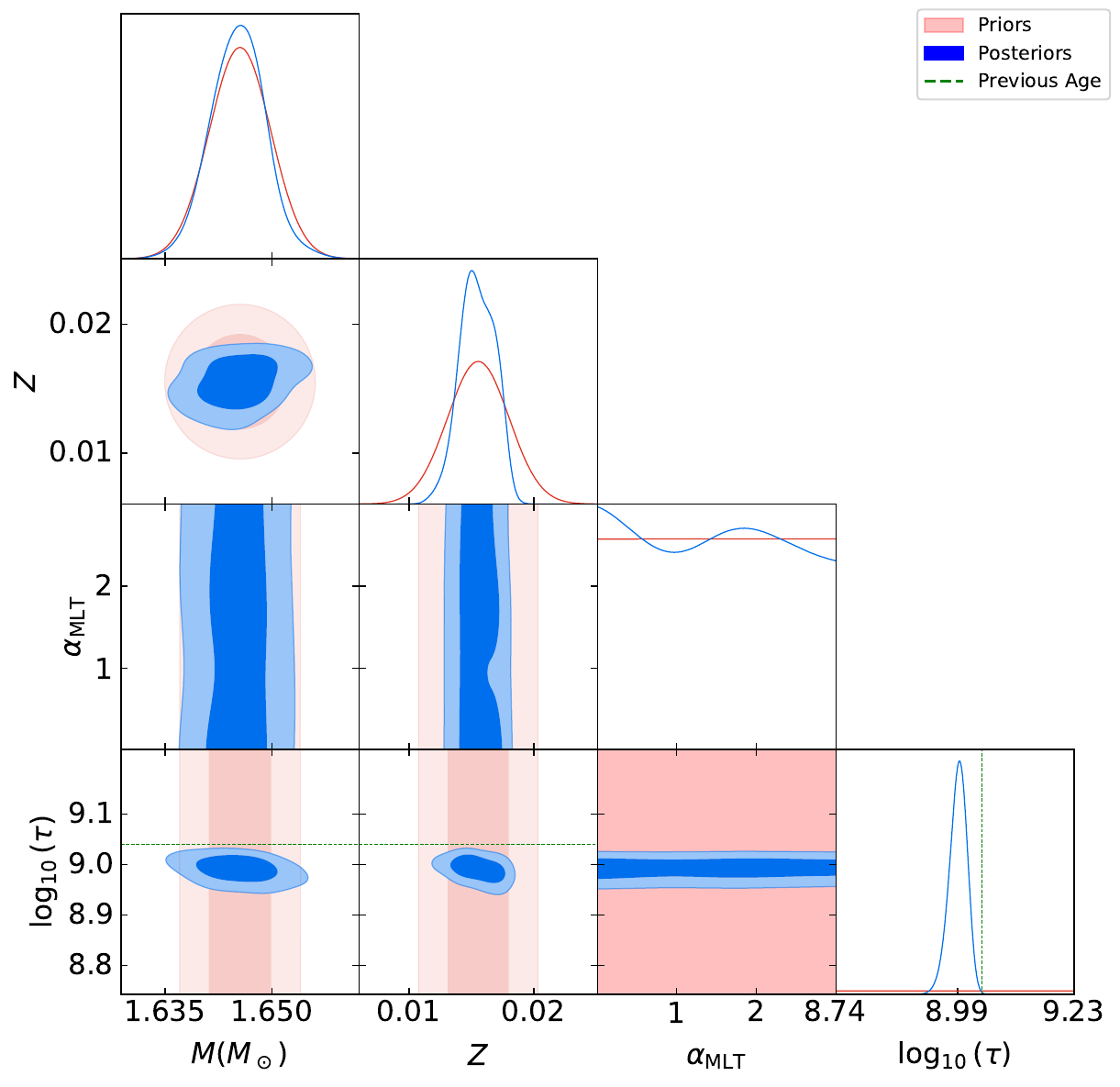}
    \caption{Corner plot showing the results of the MCMC for V501 Monoceros primary. We see that the mass is prior-dominated while our likelihood improves upon the prior inference of metallicity. Our inferred age is in agreement with the isochrone fit of 9.04 from \protect\cite{Torres2015}. $\alpha_{\rm MLT}$ is unconstrained in this case. }
    \label{fig:mon_p}
\end{figure}

\begin{figure}
    \centering
    \includegraphics[scale=0.35]{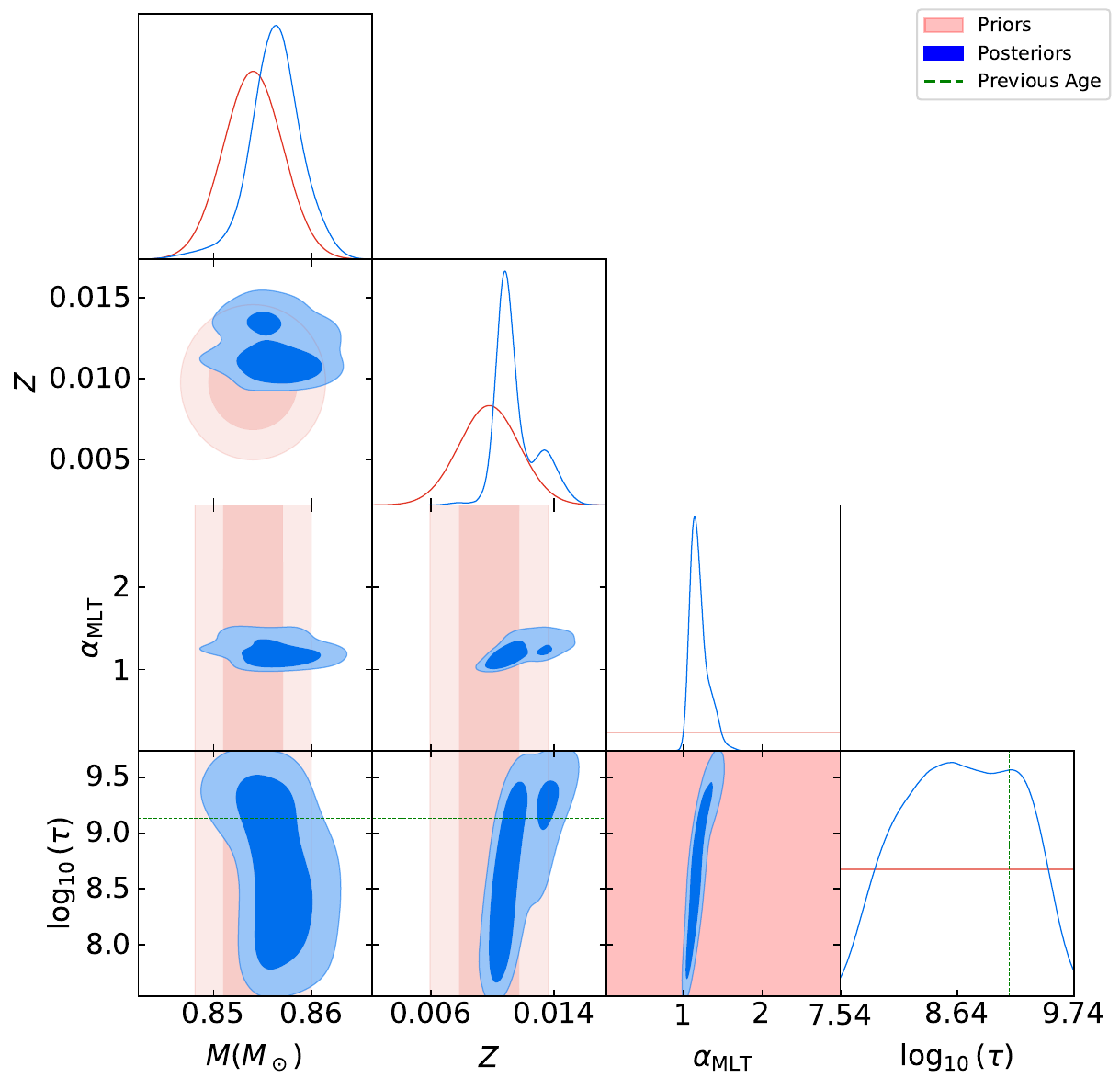}
    \caption{Same as Figure \ref{fig:mon_p} but for V636 Centaurus secondary. We infer a higher metallicity than the prior, but our posterior is consistent with it at $\sim 2\sigma$. In this case we infer values for both $\alpha_{\rm MLT}$ and {$\tau$}.
    Notably, the value for the mixing length parameter for the secondary component entirely excludes the Solar-calibrated value ($\sim 1.8-2.0$, \citet{Magic2015, Cinquegrana2022}). Interestingly, the primary component agrees with the Solar value.}
    \label{fig:v636_s}
\end{figure}

\begin{figure}
    \centering
    \includegraphics[scale=0.35]{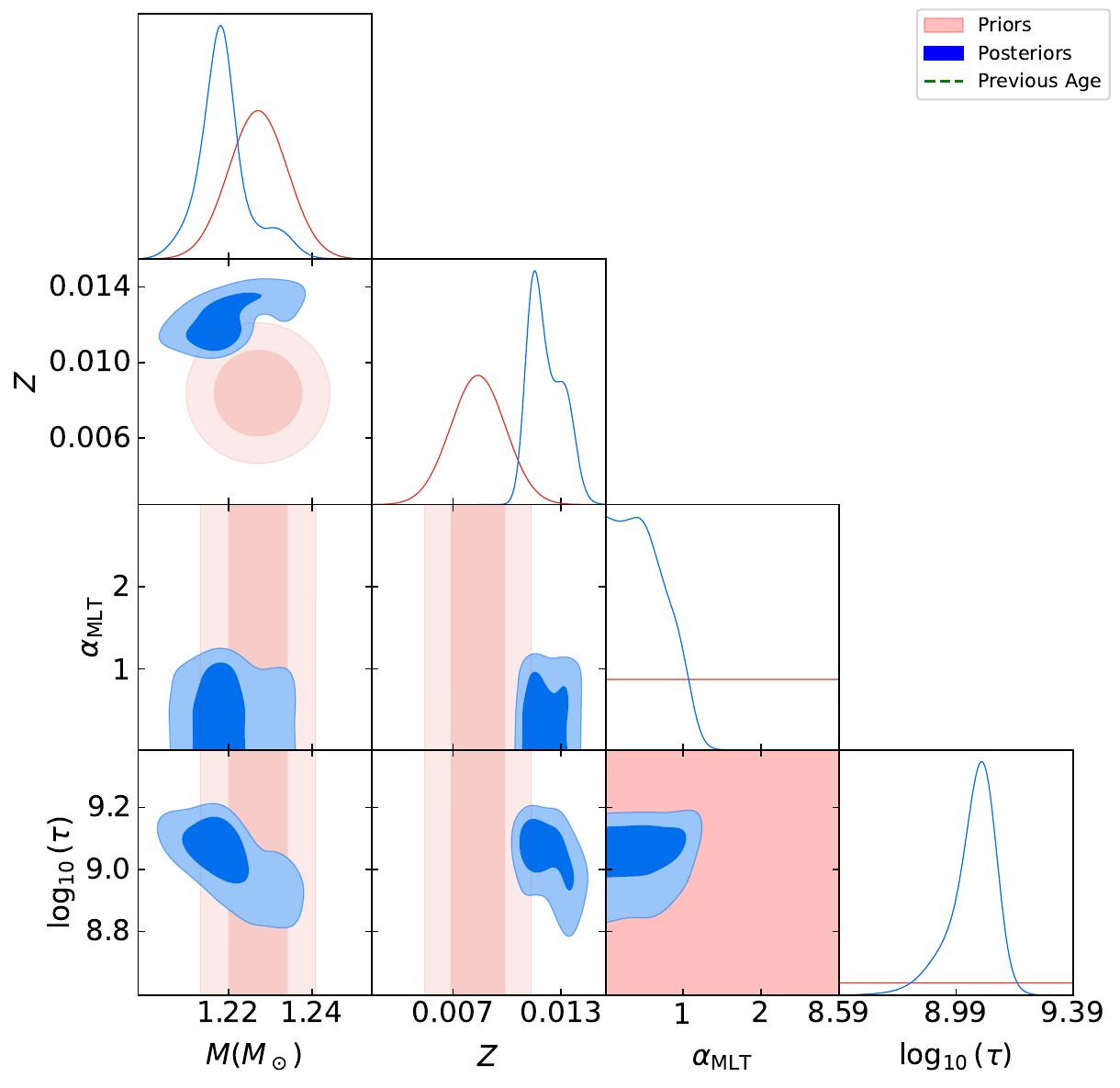}
    \caption{Same as Figure \ref{fig:mon_p} but for WZ Ophiuchus primary. As for V636 Centaurus secondary, the inferred metallicity is higher than its prior although in statistical agreement. In this case we find only an upper limit on the mixing length parameter. Previous studies using isochrone fitting e.g., \protect\cite{Clausen2008} did not find a conclusive fit for {$\tau$}, but our work constrains it to $\sim 1$ Gyr $\pm 400$ Myr.}
    \label{fig:oph_p}
\end{figure}

 We evaluated all 38 stars using separate MCMCs and found a spread of constrained values for $\alpha_{\rm MLT}$ and {$\tau$} spanning nearly the entire parameter range with all 38 successfully reaching convergence. A representative sample of some of these results are given in Figures \ref{fig:mon_p}, \ref{fig:v636_s}, and \ref{fig:oph_p}. The lower 68\% and 95\% limits from the posteriors of the $M$, $Z$, $\alpha_{\rm MLT}$, and $\log_{10}(\tau)$ parameters given by the MCMC have been collated in Table \ref{tbl:results} as either a constraint $(\pm)$ or a single bound (represented as an inequality). 

\begin{table*}
    \centering
    \resizebox{\textwidth}{!}{
    \begin{tabular}{|c|c|c|c|c|c|c|c|}
 \hline
	 Star & $M$ ($M_\odot$) & $Z$ & $\alpha_{\textrm{MLT}}$ $68\%$ [95\%] & $\log_{10}(\tau)$ $68\%$ [95\%] & Literature Age \\
 \hline
	V421 Peg p & $1.532^{+0.023}_{-0.016}$ & $0.018^{+0.002}_{-0.002}$ & --- & $8.765^{+0.134}_{-0.154}$  [8.352, 8.980] & --- \\ 
 \hline 
	V421 Peg s & $1.349^{+0.019}_{-0.017}$ & $0.012^{+0.002}_{-0.002}$ & --- & $8.274^{+0.356}_{-0.348}$  [7.717, 8.910] & --- \\ 
 \hline 
	YZ Cas s & $1.326^{+0.006}_{-0.006}$ & $0.012^{+0.002}_{-0.001}$ & --- & $8.240^{+0.431}_{-0.350}$  [7.706, 8.982] & 8.5 [\cite{Pavlovski2014}] \\ 
 \hline 
	V570 Per p & $1.425^{+0.006}_{-0.006}$ & $0.016^{+0.001}_{-0.001}$ & --- & $8.470^{+0.332}_{-0.372}$  [7.793, 9.025] & 8.78-9 [\cite{Marrese2004}] \\ 
 \hline 
	V570 Per s & $1.328^{+0.005}_{-0.005}$ & $0.016^{+0.001}_{-0.001}$ & --- & $8.418^{+0.320}_{-0.334}$  [7.849, 8.974] & 8.78-9 [\cite{Marrese2004}] \\ 
 \hline 
	V1130 Tau p & $1.302^{+0.007}_{-0.008}$ & $0.013^{+0.001}_{-0.001}$ & $< 1.368$ [2.051] & $9.166^{+0.034}_{-0.034}$  [9.101, 9.282] & 9.26-9.32 or 9.40-9.45 [\cite{Clausen2010a}] \\ 
 \hline 
	V1130 Tau s & $1.386^{+0.006}_{-0.007}$ & $0.013^{+0.001}_{-0.001}$ & --- & $9.289^{+0.036}_{-0.026}$  [9.248, 9.351] & 9.26-9.32 or 9.40-9.45 [\cite{Clausen2010a}] \\ 
 \hline 
	CD Tau p & $1.441^{+0.015}_{-0.015}$ & $0.023^{+0.003}_{-0.003}$ & --- & $9.219^{+0.075}_{-0.063}$  [9.113, 9.366] & 9.42 [\cite{Ribas1999}] \\ 
 \hline 
	CD Tau s & $1.369^{+0.014}_{-0.016}$ & $0.022^{+0.003}_{-0.003}$ & --- & $9.132^{+0.150}_{-0.126}$  [8.908, 9.416] & 9.42 [\cite{Ribas1999}] \\ 
 \hline 
	V501 Mon p & $1.645^{+0.004}_{-0.004}$ & $0.015^{+0.002}_{-0.001}$ & --- & $8.991^{+0.018}_{-0.018}$  [8.951, 9.024] & 9.04 [\cite{Torres2015}] \\ 
 \hline 
	V501 Mon s & $1.459^{+0.002}_{-0.002}$ & $0.016^{+0.001}_{-0.001}$ & --- & $8.998^{+0.053}_{-0.052}$  [8.893, 9.095] & 9.04 [\cite{Torres2015}] \\ 
 \hline 
	GX Gem p & $1.491^{+0.009}_{-0.012}$ & $0.018^{+0.001}_{-0.001}$ & $< 0.505$ [0.875] & $9.312^{+0.017}_{-0.014}$  [9.288, 9.338] & 9.45 [\cite{Lacy2008}] \\ 
 \hline 
	GX Gem s & $1.464^{+0.009}_{-0.007}$ & $0.018^{+0.001}_{-0.001}$ & $< 0.636$ [1.048] & $9.322^{+0.008}_{-0.008}$  [9.303, 9.339] & 9.45 [\cite{Lacy2008}] \\ 
 \hline 
	VZ Hya p & $1.274^{+0.008}_{-0.008}$ & $0.011^{+0.001}_{-0.001}$ & $< 1.400$ [2.201] & $8.470^{+0.324}_{-0.359}$  [7.827, 9.045] & 8.8-9.1 [\cite{Clausen2008}] \\ 
 \hline 
	VZ Hya s & $1.145^{+0.007}_{-0.006}$ & $0.011^{+0.002}_{-0.002}$ & $2.084^{+0.331}_{-0.285}$  [1.588, 2.713] & $8.441^{+0.515}_{-0.505}$  [7.697,9.217] & 8.8-9.1 [\cite{Clausen2008}] \\ 
 \hline 
	RS Cha p & $1.835^{+0.012}_{-0.013}$ & $0.019^{+0.001}_{-0.001}$ & --- & $8.904^{+0.025}_{-0.026}$  [8.846, 8.949] & $8.96 \pm 0.01$  [\cite{Kovaleva2001}] \\ 
 \hline 
	RS Cha s & $1.781^{+0.010}_{-0.010}$ & $0.016^{+0.001}_{-0.001}$ & --- & $9.009^{+0.015}_{-0.016}$  [8.977, 9.038] & $8.96 \pm 0.01$ [\cite{Kovaleva2001}] \\ 
 \hline 
	ZZ Boo p & $1.609^{+0.011}_{-0.010}$ & $0.018^{+0.001}_{-0.001}$ & --- & $9.144^{+0.026}_{-0.026}$  [9.090, 9.190] & 9 [\cite{Cester1978}] \\ 
 \hline 
	ZZ Boo s & $1.569^{+0.008}_{-0.009}$ & $0.014^{+0.001}_{-0.001}$ & --- & $9.175^{+0.020}_{-0.021}$  [9.131, 9.210] & 9 [\cite{Cester1978}] \\ 
 \hline 
	V636 Cen p & $1.052^{+0.005}_{-0.004}$ & $0.010^{+0.001}_{-0.001}$ & $1.859^{+0.199}_{-0.205}$  [1.500, 2.217] & $8.782^{+0.297}_{-0.387}$  [7.663, 9.258] & 9.13 [\cite{Clausen2009}] \\ 
 \hline 
	V636 Cen s & $0.856^{+0.002}_{-0.002}$ & $0.011^{+0.002}_{-0.001}$ & $1.193^{+0.123}_{-0.104}$  [1.025, 1.475] & $8.696^{+0.566}_{-0.597}$  [7.733, 9.587] & 9.13 [\cite{Clausen2009}] \\ 
 \hline 
	AD Boo p & $1.416^{+0.008}_{-0.008}$ & $0.021^{+0.002}_{-0.002}$ & --- & $9.094^{+0.099}_{-0.083}$  [8.959, 9.267] & 9.16-9.24 [\cite{Clausen2008}] \\ 
 \hline 
	WZ Oph p & $1.219^{+0.005}_{-0.005}$ & $0.012^{+0.001}_{-0.001}$ & $< 0.643$ [1.036] & $9.049^{+0.070}_{-0.075}$  [8.847, 9.171] & --- \\ 
 \hline 
	WZ Oph s & $1.215^{+0.005}_{-0.006}$ & $0.013^{+0.001}_{-0.001}$ & $< 0.657$ [1.013] & $9.107^{+0.058}_{-0.061}$  [8.979, 9.213] & --- \\ 
 \hline 
	V2653 Oph p & $1.567^{+0.017}_{-0.017}$ & $0.010^{+0.001}_{-0.001}$ & --- & $9.178^{+0.015}_{-0.017}$  [9.140, 9.218] & 8.3-8.6 $(300\pm100 {\rm Myr})$ [\cite{Cakirli2016}] \\ 
 \hline 
	V2653 Oph s & $1.327^{+0.019}_{-0.020}$ & $0.008^{+0.001}_{-0.001}$ & $< 0.925$ [1.463] & $9.384^{+0.014}_{-0.020}$  [9.346, 9.426] & 8.3-8.6 $(300\pm100 {\rm Myr})$ [\cite{Cakirli2016}] \\ 
 \hline 
	KIC 9851944 p & $1.624^{+0.023}_{-0.027}$ & $0.015^{+0.002}_{-0.002}$ & --- & $9.149^{+0.026}_{-0.025}$  [9.092, 9.202] & --- \\ 
 \hline 
	KIC 9851944 s & $1.998^{+0.004}_{-0.004}$ & $0.020^{+0.001}_{-0.001}$ & $< 0.918$ [2.011] & $8.958^{+0.004}_{-0.003}$  [8.953, 8.967] & --- \\ 
 \hline 
	OO Peg p & $1.790^{+0.014}_{-0.018}$ & $0.014^{+0.001}_{-0.001}$ & --- & $8.970^{+0.015}_{-0.013}$  [8.945, 8.997] & 8.84-9.11 $(1 \pm 0.3 {\rm Gyr})$ [\cite{Southworth2024}] \\ 
 \hline 
	OO Peg s & $1.727^{+0.008}_{-0.008}$ & $0.019^{+0.001}_{-0.001}$ & --- & $8.919^{+0.019}_{-0.015}$  [8.889, 8.959] & 8.84-9.11 $(1 \pm 0.3 {\rm Gyr})$ [\cite{Southworth2024}] \\ 
 \hline 
	V375 Cep p & $1.290^{+0.015}_{-0.015}$ & $0.019^{+0.002}_{-0.002}$ & $< 1.300$ [2.078] & $9.336^{+0.095}_{-0.073}$  [9.218, 9.503] & 9.53-9.59 $(3.6 \pm 0.25 {\rm Gyr})$ [\cite{Sandquist2013}] \\ 
 \hline 
	V375 Cep s & $0.873^{+0.009}_{-0.008}$ & $0.018^{+0.002}_{-0.002}$ & $1.330^{+0.178}_{-0.170}$  [1.004, 1.684] & $9.747^{+0.167}_{-0.146}$ [9.103, 10.027] & 9.53-9.59 $(3.6 \pm 0.25 {\rm Gyr})$ [\cite{Sandquist2013}] \\ 
 \hline 
	BW Aqr p & $1.464^{+0.019}_{-0.017}$ & $0.018^{+0.002}_{-0.002}$ & $< 1.456$ [2.189] & $9.280^{+0.027}_{-0.028}$  [9.224, 9.329] & 9.3 [\cite{Khaliullin1986}] \\ 
 \hline 
	BW Aqr s & $1.372^{+0.019}_{-0.020}$ & $0.016^{+0.002}_{-0.002}$ & $< 1.775$ [2.591] & $9.313^{+0.054}_{-0.052}$  [9.215, 9.409] & 9.3 [\cite{Khaliullin1986}] \\ 
 \hline 
	EF Aqr p & $1.243^{+0.008}_{-0.008}$ & $0.015^{+0.002}_{-0.002}$ & $< 1.150$ [1.514] & $8.332^{+0.385}_{-0.314}$  [7.838, 9.063] & 8.95-9.32 $(1.5 \pm 0.6 {\rm Gyr})$ [\cite{Vos2012}] \\ 
 \hline 
	EF Aqr s & $0.947^{+0.005}_{-0.006}$ & $0.015^{+0.002}_{-0.002}$ & $1.104^{+0.111}_{-0.110}$  [0.915, 1.401] & $8.719^{+0.543}_{-0.605}$  [7.712, 9.604] & 8.95-9.32 $(1.5 \pm 0.6 {\rm Gyr})$  [\cite{Vos2012}] \\ 
 \hline 
	BK Peg p & $1.413^{+0.006}_{-0.007}$ & $0.014^{+0.001}_{-0.001}$ & $< 1.317$ [1.892] & $9.316^{+0.013}_{-0.012}$  [9.294, 9.350] & 9.4-9.43 [\cite{Clausen2010b}] \\ 
 \hline 
	BK Peg s & $1.258^{+0.006}_{-0.005}$ & $0.014^{+0.001}_{-0.001}$ & $< 1.321$ [2.210] & $9.236^{+0.071}_{-0.069}$  [9.114, 9.425] & 9.4-9.43 [\cite{Clausen2010b}] \\ 
 \hline 
    \end{tabular}}
    \caption[Summary of constraints for main sequence detached eclipsing binary parameters]{The results of our analysis. A blank entry (---) indicates a posterior essentially unbounded across our parameter space. The ages found by previous works were all done using isochrone fitting. In cases where error bars in previous works were provided, we provide these in their original linear form and an accompanying [$68\%$] width in log space.~The ``Literature Age" column reports values as either a single value in log space (with an error in log space if provided), a range in log space if the reference provided a range of best fits but without reporting it as an error bar, or, if the original age was reported in linear space with an error bar, a 68\% range in log space with the original age and its error reported in parentheses. \citet{Clausen2010a} report two potential ranges for the age of V1130 Tau and we therefore provide both.}
    \label{tbl:results}
\end{table*}

\par We constrain $M$ and $Z$ in all systems.~The majority of these are within $1\sigma$ of the values in Table \ref{tbl:measurements}, and all are within $\sim 2\sigma$. We find both upper and lower bounds on $\alpha_{\rm MLT}$ for 5 stars in 4 different systems, and only one bound for another 14 stars. All but one of these 19 stars have masses $M < 1.5M_\odot$. This is the range where meaningful inferences are expected because the convective-to-radiative atmosphere transition for stars with masses $M > 1.5 M_\odot$ minimizes the effect of the mixing length parameter on the surface of the star, removing most of the constraining power on it. We obtain both upper and lower bounds on the age of all the stars in our dataset.

\par The majority of our age constraints agree well with estimates found by previous studies but, because many do not report error bounds, it is not possible to determine the precise level of agreement. A small number of our inferred ages are significantly different from previous measurements. In particular, ZZ Boo s has a discrepancy of roughly $0.17$ dex, the stars of the GX Gemini system both have discrepancies of roughly $0.13$ dex, and V2653 Oph is between $0.5$ and $1.1$ dex.

V2653 Oph has a pulsating component, which may affect our inference because the observed brightness varies over the pulsation cycle. This system is also noteworthy because it has been discussed as a likely member of the open cluster Collinder 357 \citep{Zejda2012,Cakirli2016}. Under the assumption of coeval cluster formation, its age would then be expected to be comparable to the cluster age, $\sim 60\,{\rm Myr}$ \citep{Cakirli2016}, whereas our inferred age is $\sim 1.5\,{\rm Gyr}$. We therefore regard this system as a cautionary example where unmodeled astrophysical effects, such as pulsation, may bias the inferred parameters.

\subsection{Select examples}
\par Turning to some individual examples, the results for V501 Monoceros's primary star is shown in Figure \ref{fig:mon_p}. We find good agreement with the $M$ and $Z$ priors and a near-Gaussian bound on the age in log space. The inferred age agrees with previous isochrone fitted estimates. We do not find a bound on $\alpha_{\rm MLT}$ whose posterior is essentially uniform across parameter space --- in line with our expectation above that the radiative nature of stars heavier than $1.5\msun$ will not facilitate such bounds (this star has $M \approx 1.65 M_\odot$).

\par Our results for V636 Centaurus secondary star are shown in Figure \ref{fig:v636_s}.~We infer the mixing length parameter to be $\alpha_{\rm MLT} = 1.193^{+0.123}_{-0.104}$, significantly smaller than the Solar-calibrated value. Comparing the surface gravity and effective temperatures to Solar values, V636 Centaurus secondary has a higher surface gravity and a lower effective temperature (4.53 vs 4.44 \citep{Prsa2016} for $\log(g)$ and ($5000 K$ vs $5772 K$ for $T_{\rm eff}$ \citep{Prsa2016}). However these results contradict the trend found by \citet{Magic2015} which finds as surface gravity decreases and effective temperature increases, the mixing length parameter decreases. It is worth noting that the values of $\alpha_{\rm MLT}$ \citet{Magic2015} use do not extend significantly below Solar values. We find a broad posterior for {$\tau$} that encompasses the previously fitted age. The metallicity we recover is within $\sim 2\sigma$ from the measurement provided in Table \ref{tbl:measurements}.

\par Finally, our results for WZ Ophiuchus's primary star are shown in Figure \ref{fig:oph_p}.~This is an interesting case where we have obtained constraints on all of the parameters, but the constraint for $\alpha_{\rm MLT}$ is an upper bound. The rapid decay of the posterior at $\alpha_{\rm MLT} = 1.036$ clearly limits values of $\alpha_{\rm MLT}$, but without providing a lower bound.  Once again, the central value of $Z$ does not match our prior, but the 2$\sigma$ bounds overlap significantly. We were unable to find a previous isochrone fit for this star, but we infer its age to be $\sim \textrm{1Gyr} \: \pm \sim 200 \textrm{Myr}$ using our method. 

\par All three stars in Figures \ref{fig:mon_p}, \ref{fig:v636_s}, and \ref{fig:oph_p} give tighter constraints on the metallicity than their priors. This occurs for roughly 50\% of our targets. While some of these improvements are substantial (33-50\%), there is not a uniform improvement across the data set. However, our posteriors for all of our $M$ and $Z$ constraints are within $\sim 2\sigma$ of the corresponding values in Table \ref{tbl:measurements}.

\par The complete set of corner plots for the 38 stars which converged in our sample can be found, along with the MESA model outputs, the random forest surrogates, and MESA inlists at the linked \href{https://doi.org/10.5281/zenodo.20435308}{repository} \citep{DennisDataMixing}.

\begin{figure*}
    \centering
    \includegraphics[width=\linewidth]{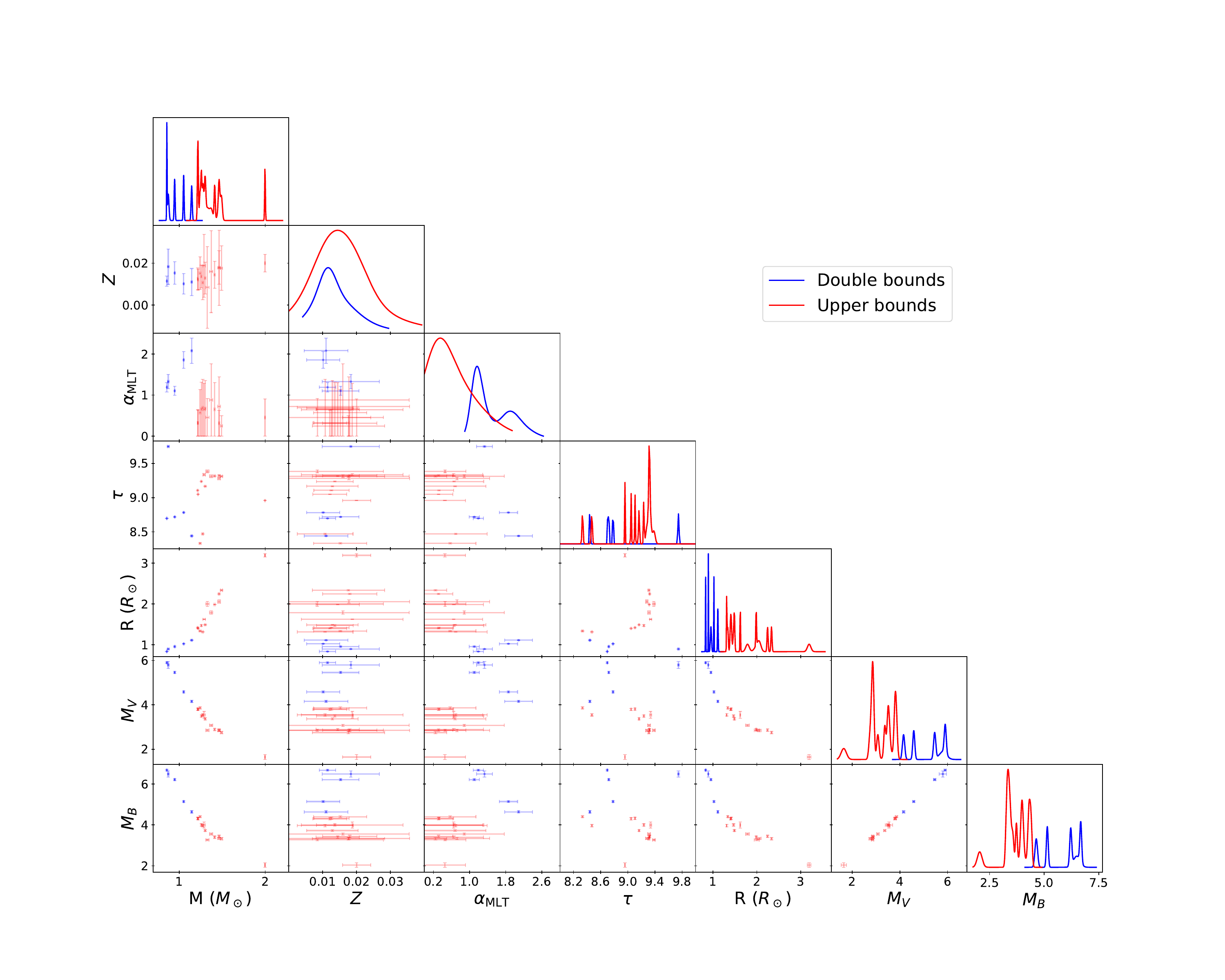}
    \caption{Corner plot of the results from our MCMC for $M$, $Z$, $\alpha_{\rm MLT}$, $\tau$ combined with the $R$, $M_V$, $M_B$ used {as input data} (Table \ref{tbl:measurements}) with scatter plots on the off-diagonal and kernel density estimates of the 1D distributions along the diagonal. The kernel width is set to the uncertainty size in each case, as in Figure \ref{fig:data_distribution}. We see a clear separation between our double sided bounds on $\alpha_{\rm MLT}$ and the upper bounded values, around around $M \sim 1.2 M_\odot$ and $R \sim 1.1 R_\odot$. }
    \label{fig:results}
\end{figure*}

Lastly, to investigate potential trends between $\alpha_{\rm MLT}$ and the other parameters, we created a corner plot (Figure \ref{fig:results}) that compiles our results similar to Figure \ref{fig:data_distribution}. From Figure \ref{fig:results} we can see that our ability to obtain two sided bounds is determined primarily by $M$ and $R$ with a clear separation between our double-bounded and upper-bounded results near $M = 1.2 M_\odot$.

\section{Discussion}
\label{sec:discussion}

\par Here we discuss our results, first considering the implications of our age and mixing length results and how they compare to previous determinations before turning to limitations of our analysis and directions for future work.

\subsection{Implications of results}
\par Our inferred ages are consistent with previous isochrone-based determinations while providing age constraints made with a fully Bayesian framework that also fully quantifies their uncertainties.~The Bayesian framework naturally captures correlations between age, mass, metallicity, and $\alpha_{\rm MLT}$, and propagates asymmetric, non-Gaussian, or multimodal posterior structure into the inferred age uncertainties.~Said uncertainties are small, $\sim 4\%$ in some cases, which are the result of the tight constraints on $M$ and $R$ provided by MS DEB stars. The precision in age appears to be anti-correlated with the age itself, with errors as large as $\sim  60\%$ in V570 Per p with a constrained age of $10^{8.5}$ years and as low as $\sim3.8\%$ in RS Cha s with a constrained age of $10^{9}$ years. {Our most precise age estimates are comparable} with other eclipsing binary star error estimates which produce relative ages with errors of roughly $\sim 5\%$ \citep{Stassun2009}. However, \cite{Stassun2009} imposed coevality by requiring both stars in each binary to share a common age, whereas we infer the age of each component independently. We also do not assume a common metallicity, although we do use the same spectroscopic metallicity prior for both components of a binary. This sacrifices some constraining power in favor of a more general model, allowing coevality to be tested a posteriori. We find that the inferred component ages agree within $\sim1\sigma$ for all but two systems, and within $\sim2\sigma$ for the remaining two, thus providing independent support for the coevality assumption. Other works have found more significant discrepancies depending on the stellar age and mass \citep{Stassun2014, Valle2016, Simon2017}, typically finding a relative agreement between independent fits to roughly 20-50\%. Our results on average fall within 30\% (e.g. CD tau $(\sim20)$, but some of our results find much better agreement (e.g. GX Gem $\sim2.3\%$ or V501 Mon $\sim 1.6\%$), {in line with the generally high precision of our inference.}

We obtain constraints on $\alpha_{\rm MLT}$ for 19 of the 38 main-sequence stars in our sample. In these systems, we infer either two-sided or one-sided limits on $\alpha_{\rm MLT}$, and in a majority of cases, the posterior distributions favor values below the Solar calibration ($\alpha_{\rm MLT} \approx$1.8–2.0 in commonly used prescriptions). While the statistical significance varies from star to star, the overall pattern is consistent with previous studies that have suggested non-universality in the mixing length parameter. Previous attempts to infer the mixing length parameter for individual stars have primarily relied on isochrone fitting or asteroseismology \citep{Noels1991, Edmonds1992, Guenther2000, Joyce2018b, Nsamba2018}. In many cases, uncertainties on the inferred $\alpha_{\rm MLT}$ values are not reported, although Bayesian isochrone techniques can in principle provide such uncertainties \citep[e.g.,][]{VallsGabaud2014}.\citet{Joyce2018a} calibrated mixing length parameters for six stars, and \citet{Joyce2018b} fit mixing length parameters and corresponding uncertainties for both $\alpha$ Centauri A and B. Of the stars constrained by \citet{Joyce2018a}, four are main-sequence stars with inferred values in the range $\alpha_\mathrm{MLT}\simeq 0.7$--$1.95$. Our findings are consistent with these results and provide independent evidence, from detached eclipsing binaries, that a single universal mixing length parameter does not describe all main-sequence stars. In another MCMC-based analysis, \citet{Desmond2021} studied Cepheid stars in DEBs in the Large Magellanic Cloud.Assuming a common mixing length parameter across their five-star sample, they inferred $\alpha_{\rm MLT}=0.90^{+0.36}_{-0.26}$.~Their individual-star fits yielded central values spanning $\alpha_{\rm MLT}\in[0.82,1.80]$, although with broad uncertainties, indicating that the data preferred generally sub-Solar values but did not tightly resolve system-to-system variation.

Allowing $\alpha_{\rm MLT}$ to vary has direct implications for age determination. In traditional isochrone fitting, the mixing length parameter is fixed and age is inferred conditional on that assumption. Our analysis demonstrates that age and $\alpha_{\rm MLT}$ can be partially degenerate on the main sequence, but that the combination of precise radii and photometric constraints can nonetheless restrict the allowed parameter space. For most systems, the inferred ages are consistent within uncertainties with previous determinations, suggesting that fixing $\alpha_{\rm MLT}$ does not generally induce catastrophic biases in these cases. 

\par Modern observations have enabled empirical estimates of the mixing length parameter in other systems such as dwarf and red giant branch stars \citep{Bonaca2012, Tayar2017, Joyce2018a}.~Despite this progress, these studies do not yet agree on either a universal value or a simple scaling relation for $\alpha_{\rm MLT}$ across stellar parameters \citep{Joyce2023}. The results presented here therefore provide additional empirical support for the growing evidence that the mixing length parameter may not be universal, using main-sequence detached eclipsing binaries as an independent probe.

\par Regarding $M$ and $Z$, our work does not yield significant improvements upon previous studies. These parameters have tight priors from other measurements and our MCMC for the most part recovers those. Notably however, in the case of V2653 Oph s, the width of constraint in $Z$ is roughly 40\% the original (down to roughly $\pm0.001$ from $\pm0.004$), and there are several other stars for which the constraint on metallicity is considerably narrower than the prior. Importantly in both $M$ and $Z$, despite some values being more precisely or less precisely constrained there are no cases of significant disagreement ($\gtrsim 2\sigma$) between the {values in Tables \ref{tbl:measurements} and \ref{tbl:results}}.

\par Beyond the astrophysical interpretation, our work also expands the field of techniques \citep{Jorgensen2005, Barnes2007, Maxted2015, Lebreton2020} which have been developed to replace visual isochrone fitting. Even these approaches have typically relied upon fitting interpolated grids, whereas ours improves upon them by developing a method that allows for a more efficient sampling of parameter space while simultaneously avoiding the limitations of traditional interpolation by using machine learning. Our results demonstrate the feasibility of combining active learning with machine learning emulation to perform fully Bayesian inference with detailed stellar evolution models. Direct MCMC analyses calling MESA at each likelihood evaluation are computationally expensive, particularly when exploring multi-dimensional parameter spaces. By concentrating training models in regions where the surrogate is uncertain, the active learning approach reduces the number of required stellar evolution runs while maintaining accuracy at the level of observational errors. The framework is also highly scalable: additional input parameters (e.g., overshooting, diffusion efficiencies, or alternative abundance prescriptions) could be incorporated in future work, and larger stellar samples may readily be accommodated. As stellar datasets continue to expand in size and precision, such approaches will be increasingly important for systematically confronting stellar evolution theory with data.

\subsection{Limitations}

\par To our knowledge, this is the first study to combine machine-learning, active-learning, and MCMC to constrain stellar physics using MS DEBs, so there are necessarily limitations. The most important is that we only vary $M$, $Z$, and $\alpha_{\rm MLT}$. Ideally, other parameters, particularly $Y$, opacity, differences in composition, and nuclear reaction rates which can affect the effective temperature and stellar lifetime should also be varied. Additionally, we only employ a single stellar structure code, which may introduce systematic differences relative to other codes. Lastly, our sample only contains 38 stars in a specific phase of evolution. This is hardly a conclusive sample and does not account for changes to the mixing length that may occur for different stellar types and phases of stellar evolution.

\subsection{Future Work}
\par Future studies can address these limitations by extending the inference to additional stellar physics parameters that are fixed in the present work. This is particularly important for systems with $M \gtrsim 1.5 M_\odot$, where the envelopes are predominantly radiative and $\alpha_{\rm MLT}$ has little effect on the observables considered here. Such systems may instead provide stronger constraints on other aspects of the stellar models. However, as we already have no bounds on  for a large number of systems, it may not be possible to constrain additional parameters. This motivates follow-up studies with additional bands such as the $U$ or $I$ bands which may provide more information, particularly at the edges of our mass range. Additionally, we could use our machine learning emulator in a Bayesian hierarchical model to determine whether $\alpha_{\rm MLT}$ systematically varies with other stellar parameters.

Further, there are other areas where the need to run large training grids can be overcome with the active learning technique developed in this work. The use of machine learning surrogates as replacements for expensive modeling codes is becoming commonplace in astronomy and astrophysics \citep[e.g.,][]{Conceicao2024, Maltsev2024, Hirashima2025, Tahseen2025, Vermarien2025}. However, these rely on large numbers of precomputed model grids, which are computationally prohibitive. The active learning method developed here demonstrates the effectiveness of creating a targeted training set where the ML surrogate does not need to predict all of parameter space accurately, only the relevant areas.

Finally, we treat each binary component independently and do not enforce co-evolution (i.e. a shared age and metallicity). This makes our approach conservative but discards a well-motivated physical prior.~{Since our results support coevolution, our constraints can likely be safely tightened by enforcing it.}

\section{Conclusions}
\label{sec:conclusion}
{We have presented a Bayesian framework for inferring the mass ($M$), metallicity ($Z$), mixing length parameter ($\alpha_{\rm MLT}$), and age ($\tau$) of main-sequence detached eclipsing binary stars (MS DEBs). The method combines Gaussian priors on the measured masses and spectroscopic metallicities with likelihoods for the observed radii and broadband magnitudes, and uses MCMC sampling to propagate degeneracies between parameters and obtain statistically rigorous uncertainties. To make this inference computationally feasible, we trained machine-learning surrogates for MESA stellar evolution models using an active learning procedure that adaptively selected new models in the regions of parameter space most relevant for improving emulator accuracy. Our final training set contains $\sim 20,000$ stellar models, corresponding to more than one million points along their evolutionary tracks.

Applying this framework to 38 stars in MS DEBs, we obtain age constraints for every star in the sample and bounds on $\alpha_{\rm MLT}$ for a subset of systems.~The age constraints are broadly consistent with previous isochrone-based determinations while providing full posterior uncertainties and parameter correlations within a Bayesian framework.

For the mixing length parameter, we find double-sided constraints for 5 extra-Solar main-sequence stars and one-sided bounds for an additional 14, using only observations of $M$, $Z$, $R$, $M_V$, and $M_B$. The constraints occur primarily in lower-mass stars, where convective envelopes make the observables sensitive to $\alpha_{\rm MLT}$, while hotter and more radiative stars generally provide little constraining power on this parameter. This pattern provides a useful physical consistency check on the inference.~Several of the inferred mixing length parameter values differ from the commonly adopted Solar-calibrated value of $\alpha_{\rm MLT}\simeq 1.8$--$2.0$, supporting previous findings that a universal mixing length parameter may not adequately describe convection across the main sequence. 

Only the parameters above were varied, with all other stellar inputs held fixed. Future studies can extend the framework by varying additional parameters, such as helium abundance, diffusion, overshooting, and atmosphere/boundary conditions, thereby quantifying the associated systematics.

As precise stellar measurements become more abundant, frameworks such as ours provide a scalable way to connect those data to detailed stellar evolution models. By combining active learning emulators with Bayesian inference, this approach can be extended to additional stellar physics parameters and applied to other well-characterized stellar systems, enabling both improved stellar parameter estimates and tests of the physical assumptions entering stellar evolution models.}

\section*{Acknowledgments}
We are grateful for conversations with Peter Sadowski and David Rubin.~We are especially thankful to David Schanzenbach for his assistance with using the University of Hawai\okina i KOA supercomputer.
HD and JS thank Bhuvnesh Jain and The University of Pennsylvania for their hospitality while part of this work was completed. JS thanks ICG Portsmouth for hospitality and acknowledges an IPPP DIVA fellowship to support the visit.

MD is partially supported by the University of Hawai\okina i at M\=anoa Graduate Student Organization Award: 26-1-01.

HD is supported by a Royal Society University Research Fellowship (grant no. 211046).

The technical support and advanced computing resources from University of Hawaii Information Technology Services - Research Cyberinfrastructure, funded in part by the National Science Foundation CC* awards \#2201428 and \#2232862 are gratefully acknowledged.

%\section*{Software}

%%%%%%%%%%%%%%%%%%%%%%%%%%%%%%%%%%%%%%%%%%%%%%%%%%
\section*{Data Availability}

All of our data can be found at the Zenodo repository linked \href{https://doi.org/10.5281/zenodo.20435308}{here} \citep{DennisDataMixing}.



\bibliographystyle{mnras}
\bibliography{example} % if your bibtex file is called example.bib

%\begin{thebibliography}{99}
%\end{thebibliography}

%%%%%%%%%%%%%%%%%%%%%%%%%%%%%%%%%%%%%%%%%%%%%%%%%%

%%%%%%%%%%%%%%%%% APPENDICES %%%%%%%%%%%%%%%%%%%%%

%\appendix

%\section{Results for select extra targets}


%%%%%%%%%%%%%%%%%%%%%%%%%%%%%%%%%%%%%%%%%%%%%%%%%%

% Don't change these lines
\bsp	% typesetting comment
\label{lastpage}
\end{document}